# AGENT-BASED DECISION MAKING FOR INTEGRATED AIR DEFENSE SYSTEM

## Sumanta K. Das and Sumant Mukherjee[1]

**Abstract.** This paper presents algorithms of decision making agents for an integrated air defense (IAD) system. The advantage of using agent based over conventional decision making system is its ability to automatically detect and track targets and if required allocate weapons to neutralize threat in an integrated mode. Such approach is particularly useful for futuristic network centric warfare. Two agents are presented here that perform the basic decisions making tasks of command and control (C2) like detection and action against jamming, threat assessment and weapons allocation, etc. The belief-desire-intension (BDI) architectures stay behind the building blocks of these agents. These agents decide their actions by meta level plan reasoning process. The proposed agent based IAD system runs without any manual inputs, and represents a state of art model for C2 autonomy.

## INTRODUCTION

Conventional ways of decision making for command and control (C2) for an Integrated Air Defense (IAD) system are performed by human decision makers. The term IAD means that different tactical air defense services like searching, detecting, tracking, identifying and engaging targets using air defense sensors (radars) & weapons (aircrafts & missiles) are performed in an integrated fashion. Network Centric Warfare (NCW) is a concept that makes IAD operations successful. The C2 of NCW is viewed as a collaborative decision making process. With the advent of synchronous or asynchronous NCW in terms of both time and space [1], the conventional methods and modes of implementing the decision making processes of C2 has become obsolete [2]. The modern networked-laid IAD systems demand advanced method of decision making that should be enriched with artificial intelligence (AI) techniques. At each level of service execution decisions need to be taken autonomously by intelligent computational entities or agents. These agents should be capable to take localized decision and communicate with each other to achieve a collective goal.

The Belief-Desire-Intension (BDI) architectures [3] are based on the philosophical tradition of understanding practical reasoning. Recently these architectures are extended to develop autonomous agents on the basis of a number of disciplines ranging from the economics to cognitive psychology to mathematics. The BDI architectures are applied for developing agents that behave deliberatively and reactively in a complex environment. In these architectures, the mental attitudes of the agent are represented by the attributes like beliefs, desires and intentions. The belief is the knowledge of the agent about its world or environment. Agent's desire or goal is the condition that agent wants to satisfy. After satisfying the conditions, the agent has to perform certain action to achieve the goal that is known as intentions. Agents have different course of actions to achieve different desires or goals i.e. known as plan repository or plan library. JACK is the most widely used programming language for developing the BDI agents [4].

In the recent times several studies have been performed to understand and improve the agent based modeling in different application domains. The agent technologies have been successfully deployed for wireless battery powered sensor network in [5] for graph colouring problem. Agent based modeling and simulation tools are used for making automated car driver [6]. Software agents can be embedded on the web as a replacement of the human user. These agents can do the work which human user is supposed to do. In such situation dynamic service composition is essential. In [7] a work is presented where agents are evolving service semantics cooperatively in a consumer driven approach. An application of distributed computation by multi agent system for traffic control is presented in [8]. Introducing learning capability in BDI architectures is studied by [9]. A new architecture is presented in that study as an extension of the BDI architectures in which learning process is described as plans. The manipulative abduction that reasons by experiences and exhibitions of behaviour to find some pattern in the environment is used for the learning process.

Search is an essential part of the agent's model. It is a sequence of actions that takes any agent from the initial state to the goal state. Search could be uninformed or informed (heuristic). Heuristic search is an essential action for agents that work in the real time. Two classes of heuristic search methods are common, namely real time heuristic search and incremental heuristic search. A detail comparison of these two methods along with their advantages and disadvantages is presented in [10].

The agent models are difficult to verify because there is always a gap of understanding between agent logic and agent programming. To overcome this problem an operational semantics of agent programming language is presented in [11]. In that study agent logic is first grounded by state based semantics then denotational semantics is used to connect the agent logic with agent programming.

An agent may pursue multiple goals at same time. In such situation it may happens that pursuing multiple goals at the same time simultaneously is not possible. This is known as conflicting goals situation. The semantic representation of conflicting goals is presented in [12]. Monitoring many agents in a multi-agent architecture is a viable problem of agent development work. Usual disagreements between different agents arise in such situation.

In the present study C2 agents that are capable of taking autonomous decisions are identified, designed and implemented for an IAD system. The OODA approach [13] (i.e. Observe-orient-decide-act) is assumed for modeling the tactical behavior of these agents. This control loop has since long been used for understanding the human participation in the complex C2 problem. Major roles of C2 of air defense systems are threat assessment (TA) and weapon allocation (WA). The objective of this study is to apply the practical reasoning process of human decision makers to develop autonomous agents responsible for TA and WA. The BDI architectures are most suitable for implementing the philosophical tradition of understanding practical reasoning. These architectures are also suitable for developing team of agents as similar to the hierarchical structure of air defense system.

---

[1]　Institute for Systems Studies and Analyses, Defence Research and Development Organization, Metcalfe House, Delhi-110054, India.

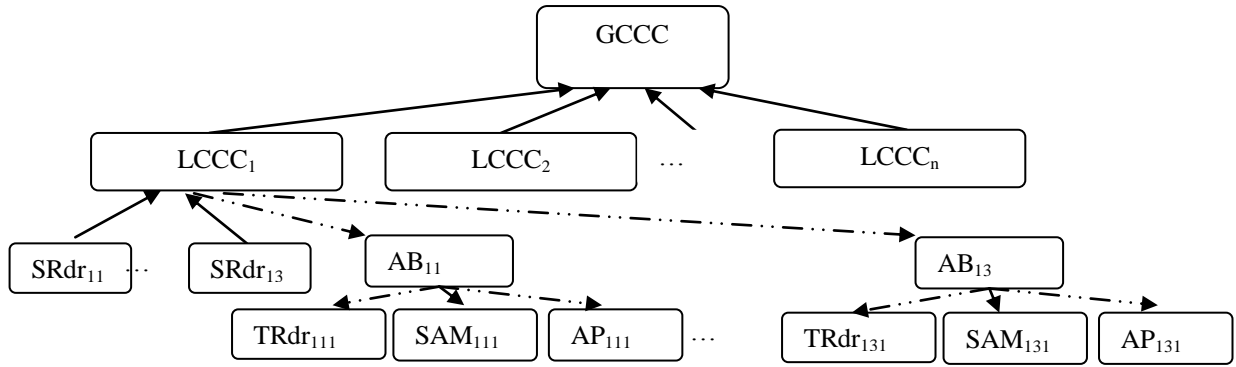

**Figure 1.** Hierarchical structure of a standard IAD system.

This paper is intended to contribute the application of BDI architectures, as an extension of the goal based agent architecture, for developing the decision making agents for an IAD system. Main concern is to formulate the mental attributes of a human decision maker in terms of belief, desire and intension. This is a novel methodological application of agent based modelling for IAD system and a technology integration between agent oriented programming and Java based combat simulation model. Two decision making agents are proposed. The first one is related with identification of jamming by surveillance radar in battlefield and the second one is related with TA and WA. A brief discussion is presented about the deployment status of these agents in a simulated air combat scenario along with the lessons learned from this study.

## IAD SYSTEM

Air defense system has progressed steadily over recent years to include highly sophisticated mission planning tools and artificially intelligent capabilities [14, 15]. The Air force mission support system (AFMSS, [14]), Power Scene and Top Scene [14] all represent major advances in this field. Up to now, most IAD operations consist of large teams of human operators that control the IAD's actions.

The idea of using multi agent system (MAS) for weapons and targets management in IAD is appropriate for distributed architectures. In cooperative MAS, agents work together to achieve one or more desired common goals. The overall system goal is achieved through interactions and coordination of the individual agents [3]. A distributed agent team has advantages over a single, complex agent in many applications [16]. For example, for search and rescue operations, multiple robots can forage far more effectively than a single, complex robot [17].

## COMMAND AND CONTROL OF IAD SYSTEM

This section is intended to identify the possible information processing agents for performing the task of C2 in an IAD system. C2 of air defense system of most of the countries follow certain hierarchical structure. Information is being exchanged between different levels of this structure. The Global Command and Control Centre (GCCC), Local Command and Control Centre (LCCC), Surveillance Radar (SRdr), Airbase Cadre (AB), tracking radar (TRdr), surface to air missile (SAM), aircraft pilot (AP) are the different components of an IAD system. The main roles of these components are target detection and classification, threat assessment (TA) and weapons allocation (WA).

Figure 1 shows the hierarchical structure of a standard IAD system. The directions of flow of information between different levels are shown by arrows. The arrow with dashed line is purposely used to represent that command is passing from higher to lower level. At the top of the tree is the higher command unit which is known as the Global Command And Control Centre (GCCC). Therefore in a MAS set-up these are identified as one agent; namely the GCCC agent.

The GCCC agent first analyses the decisions given by the different LCCCs located at diverse locations and takes its own decision then passes it to the next level of the command units i.e. to AB. The LCCC unit is identified as the second agent. This agent analyses the information given by the different SRdr(s) at diverse locations and takes their own decision based on its perception and passes it to the GCCC agent. The SRdr is the third type of agent. The LCCC agents decide which target to engage and which weapon to allocate to that target. The AB is identified as the fourth type of agents. Based on the decisions given by the LCCC agent, the AB agent decides which TRdr to track which target and which SAM to engage which target. The TRdr and the SAM systems are considered to be the fifth type of agents. Based on the decisions given by the AB agent these agents engage targets. In this study only two agents (namely, SRdr and LCCC agent) are designed and implemented to show the paradigm shift of agent based decision making for an IAD system.

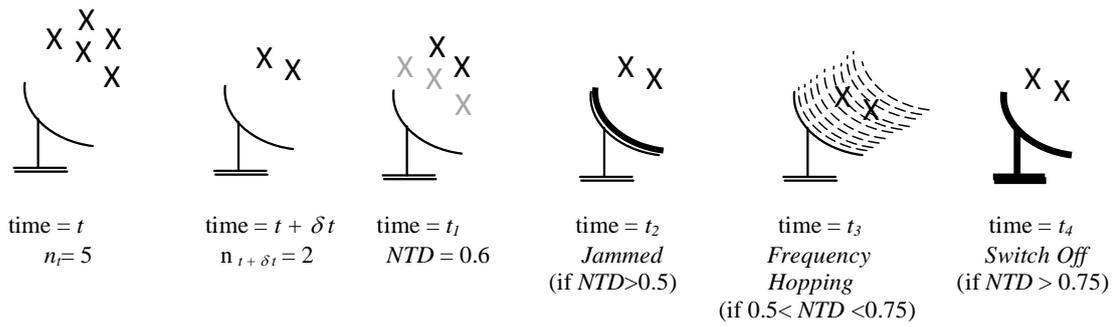

**Figure 2. Working principle of surveillance radar agent, $n_t$ stands for number of targets detected at time $t$. X denotes target detected and X denotes the hidden target.**

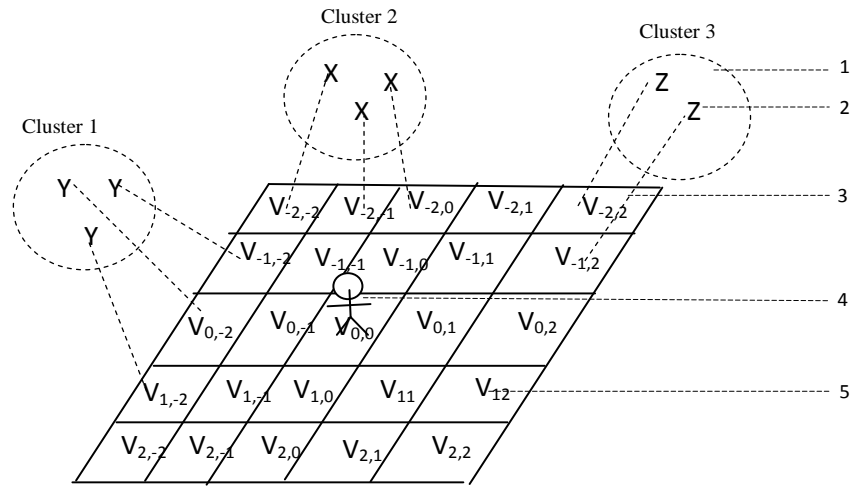

**Figure 3. LCCC agent (4), prioritizing the clusters (1) and performing autonomously the target-interceptor pairing (action) (3). Different targets (2) are grouped in different clusters (1) by MSDF module. For prioritizing the clusters the LCCC agent uses the *VAVP* values (5).**

## BDI ARCHITECTURES OF C2 AGENTS

Two main questions are answered while constructing the BDI architectures of the C2 agents. First one is what goals (options or desires) the agent decides to achieve with its current beliefs about the environment and second is how it is going to achieve these chosen goals (intensions) by means of some actions. These issues are resolved from the practical reasoning applied by the human experts to the air defense domain.

### A. BDI architectures of the surveillance radar agent

Surveillance radars are required to detect aircrafts or missiles flying towards them and often misdirected or confused by the enemy targets that uses noise jamming. The experienced radar operators can detect jamming and they generally decide to keep the radar switch off in such situation. The goal of the SRdr agent is to protect the radar from the noise jamming. From the intensity of jamming this agent can decide which action will be suitable for the radar.

The SRdr agent measures the intensity of jamming from the difference between the numbers of targets detected at time $t+1$ and $t$. If the difference is significant, the radar is jammed. The working principal of SRdr agent is shown in the Figure 2. The SRdr agent is assumed to be deployed in the simulated environment. It receives the information like "numbers of targets ($n_t$)" detected at time $t$ from its environment. On the basis of $n_{t+1}$ and $n_t$ it identifies the occurrence of noise jamming. An index is defined for this purpose namely Normalized Target Difference ($NTD = |\ ((n_t-n_{t+1})/n_t)\ |$). On the basis of the *NTD* values it decides what action it should perform based on its belief. The main action of the SRdr agent is to perform target detection in a jamming free environment. Depending on the *NTD* values, the SRdr agent can stay either in any two of the four states namely "*Sense Mode*", "*Sleep Mode*", "*Switch Off*" and "*Frequency Hopping*". If the *"Jamming"* is found then it can go either in "*Switch Off*" or in "*Frequency Hopping*" mode.

### B. BDI architectures of the LCCC agent

**Input:** cluster id, package size, mission type, cluster locations, number or attacking aircraft. *VAVP* locations, interceptor locations.
**Output:** target priority list, target-interceptor pairing.

1. Initialize *clock* = 0, *simulation_time* = 60, LCCC agent *A*;
2. **While** ( *clock < simulation_time*)
3.    *A.distance ()* ;
4.    *Clock* ++ ;
5. **Endwhile**
   **3.** *A.distance ()*
   **Start:**
   3.1. Compute distances ($d_1$, $d_2$) between clusters and VAVP and clusters and interceptors;
   3.2. Add the distances in a *beliefset-1* and *beliefset-2* respectively.
   3.3. Add attacking aircraft ranking in a *beliefset-3*.
   3.4. Add attacking aircraft and interceptor availability separately in *beliefsets-4* and *5* respectively;
   3.5. Post an event ($ev_1$) confirming that all belief updating is complete;
   3.6. Meta-level plan reasoning using *beliefset-1* to find closest clusters;
   3.7. Post the closet cluster information by an event $ev_2$.
   3.8. Meta level plan reasoning using *beliefsets-2* and *5* to find closet interceptor.
   3.9. Post the cluster-interceptor pairing with an event $ev_3$.
   3.10. Meta level plan reasoning using *beliefsets-3* and *4* to find closet attacking aircraft in the cluster.
   3.11. Update the *beliefsets-4* and *5*.
   **End**

**Figure 4. Algorithm for implementing the LCCC agent.**

**Input:** number of targets (*n*) at *t* and *t+1*.
**Output:** identifying jamming, and action against jamming.

1. Initialize *clock* = 0, *simulation_time* = 60, Surveillance Radar Agent *S*;
2. **While** ( *clock < simulation_time*)
3.    *S.update (clock)* ;
4.    *Clock* ++ ;
5. **Endwhile**
   3. *S.update (clock)*
   **Start :**
   3.1. Calculate *NTD* as ratio of $n_t / n_{t-1}$.
   3.2. Add the *NTD* and clock time in *beliefset-1*.
   3.3. Automatically post an event ($ev_1$) containing *NTD value* and *clock* time.
   3.4. Meta level plan reasoning (*plan-1*) for no action if *NTD* value is less than 0.5.
   3.5. Meta level plan reasoning for action against jamming (*plan-2*) if *NTD* value is greater than 0.5.
   3.6. Posting an event $ev_2$ from *plan-2* containing *NTD value* and *clock* time.
   3.7. Meta level plan reasoning for frequency hopping action (*plan-3*) if *NTD* value lies between 0.5 to 0.75.
   3.8. Meta level plan reasoning to switch off the radar (*plan-4*) if *NTD* value is greater than 0.75.
   **End**

**Figure 5. Algorithm for implementing the surveillance radar agent.**

The LCCC agent is responsible for TA and WA. This agent gets inputs from the multisensory data fusion (MSDF) module. The MSDF groups the detected targets into different clusters and sends the cluster information (cluster identity, cluster location) along with the situational assessed inputs about the enemy's intent (like mission type i.e. strike or escort, package size i.e. small or large). Based on this information and the LCCC agent's own beliefs (like *VAVP* (vulnerable area and vulnerable points) value), LCCC agent prioritizes the clusters and allocate interceptor to the attacking aircraft. Figure 3 shows the LCCC agent residing in a grid environment, evaluating the threat and prioritizing targets along with target-interceptor pairing. For finding the closest cluster this agent uses the meta-level plan reasoning (MLPR) process based on the distance measure. The goal of the LCCC agent is to optimally engage the detected targets with its available interceptors subject to the restriction that no target gets engaged by more than one interceptor.

## META LEVEL PLAN REASONING

In this study, the concept of MLPR [3, 4, 9] is used extensively by the C2 agents for taking optimal decisions. MLPR is a method of selecting the appropriate plan from the plan library to satisfy the agent's goal. This method is generally used for BDI agent implementation. The actions in MLPR are supposed to be optimal in some respect. Sometimes, MLPR can also be used to enable the agent to learn from the changing environment.

MLPR is implemented by using the *getInstanceInfo()* library function provided by the JACK [17]. The *getInstanceInfo()* method calculates the ranking of a plan by a *PlanInstanceInfo* object. The ranking is done by calculating one index which is a function of distance, mission type and package size. Mission types and package sizes are assumed to be fuzzy set. The membership values of these variables are obtained by using a trapezoidal membership function. Each distance, package size and mission type generates a distinct plan (Figure 7). The plan with maximum ranking get selected by the *getInstanceInfo()* function. In this way MLPR capability is incorporated in the LCCC agent's architectures. The events are posted either by the LCCC agent itself or by other plans. For example, the event *ev2* (namely *NewClusterPriorityEvents*) is posted from the plan "*NewClusterPlan*" when this plan is selected by the agent. In the similar way the interceptor aircraft of the defender force is allocated to the nearest aircraft. While allocating an interceptor to the aircraft the agent also checks its availability status so that multiple allocations do not take place. The agent's algorithms are shown in the Figures 4 and 5 and implemented through the JACK agent programming language as shown in the Figure 6 (a and b).

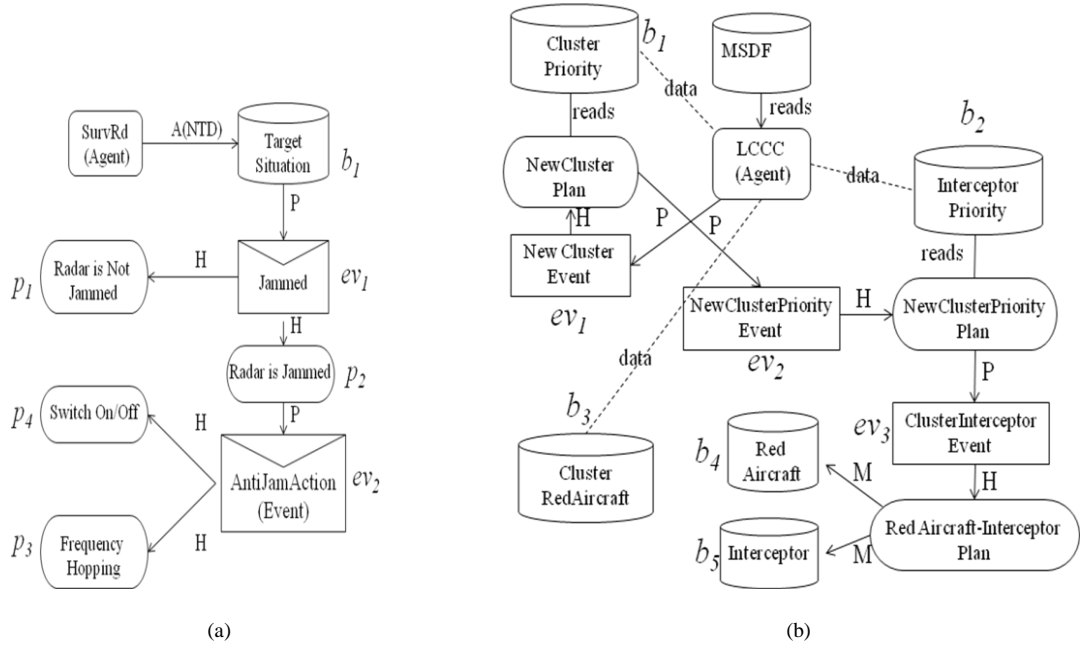

**Figure. 6.** Architecture of (a) surveillance radar agent and (b) LCCC agents developed through JACK agent programming language. Abbreviations: *b*: beleifset, *ev*: event, *p*: plan, *H*: handles, *P*: posts, *M*:modifies, *A*: add, *NTD*: Normalized target difference.

The agent can decide which of the plans are applicable for a particular event using either the three functions separately or together. First one is the *relevant()* function. The agent uses this function to select the plans that can handle such particular event. Second is the *context()* function. This function is used to select plan instances which are consistent with the agent's current beliefs. If there are still multiple plans left in the applicable plan set, the JACK provides the *getInstanceInfo()* function to return a *PlanInstanceInfo* object. This class has the *def()* method which can return the rank of the plan. The MLPRs of the "*NewClusterPlan*" of the LCCC agent and "*Radar is Jammed*" plan of the SRdr agent are shown in the Figure 7. The rank of the "*NewClusterPlan*" plan instance is the function of normalized distance between the cluster mean and the *VAVP* value, package size and mission type. The SRdr agent uses the *relevant()* function for MLPR where as the LCCC agent uses both the *context()* and the *getInstanceInfo()* functions together.

**EVALUATION**

Two approaches are used for evaluating the C2 agents. First approach is the logical evaluation and second approach is the statistical evaluation. Although logical evaluation is the most widely used method for agent research, it can not quantify the performance of the agents. A solution could be to use logical evaluation for identifying the deadlock situations and quantifying the performance by statistical measures. The logical verification rectifies the conflicting/multiple goal situations in the system. The statistical hypothesis testing measures the performance of the agents.

*A. Logical Evaluation*

First approach is based on the logical verification of the agent's model. In logical verification, the concept of goal inference rule (gir) is used for detecting the conflicting goals in the system ([12]). The girs' for SRdr agents is defined as in the Figure 8. For example the gir,

$$\{Jammed\}^{\beta}, \{Frequency\ Hopping\}^{k-} \Rightarrow Switch\ Off$$

represents that if the SRdr agent is *Jammed* (belief state denoted by $\beta$), it may derive the goal to go for the "*Switch Off*" plan (denoted by $\pi_1$), but the goal to go for the "*Frequency Hopping*" (denoted by $\pi_2$) is in conflict (denoted by $k-$, where as $k+$ denotes non-conflicting goals) with the goal to go for the "*Switch off*" plan (see Figure 8).

The girs are extended to default logic ([12]). The gir helps to find out any sort of conflicting goals present in the model. Consider that one wants to express that if a SRdr agent is found "*Jammed*", it may go either for the "*Switch Off*" or "*Frequency Hopping*" mode, but should not simultaneously pursue these goals simultaneously, i.e., the goals "*Switch Off*"

```
context()
   {
       ClusterPriority.get ($cls_id, $pck_size, $msn_type, $vavp_id, $distance);
   }
```

```
static boolean relevant(Antijam ev)
   {
       return (ev.NTD >0.75);
   }
```

```
public PlanInstanceInfo getInstanceInfo()
{
        try {
        if ($pck_size.equals("Big")) value1 = 2; else  value1 = 1;
        if ($msn_type. Equals("Strike")) value2 = 2; else value2 = 1;
        rank = (int) ($distance.as_double() / 100.0 + value1 / 2 + value2 / 2);
        return PlanInstanceInfo.def [9-rank];  }
        catch (LogicException ex)  {   return PlanInstanceInfo.def [0]; }
}
```

**Figure 7** Meta level plan reasoning by *context()*, *relevant()*, and *getInstanceInfo()* functions. The *relevant()* method in this diagram is part of the "Radar is Jammed" plan of the surveillance radar agent. The *context()* and *getInstanceInfo()* functions are part of the "*NewClusterPlan*" plan of the LCCC agent.

and "*Frequency Hopping*" are conflicting. Moreover, if a SRdr agent has "*Switch Off*" mode, it wants to go "*Sleep Mode*" with it, if it is in the "*Frequency Hopping*" mode, it wants to remain in the "*Sense Mode*". This could be modeled using the girs, as shown in the Figure 8.

Another form of logical evaluation adopted in the study is representing the entire mechanism in the form of operational semantics [11]. In this approach first agent model is grounded with state-based semantic, then denotational semantics are used to define the mathematical relation connecting agent logic and agent programming. The operational semantic, state based semantic, model semantic of LCCC agent use a propositional language ($L_0$) to represent their environment with the operators like ∧ (conjunction), ∨ (disjunction) and ¬ (negation). The $L_0$ is infinite set of atomic proposition that uses entailment ( |= ) relation. The operational semantic defines the input-output relation as a compositional function mapping from initial states to the final state reached upon termination. The state based semantics provides the ingredients for defining the operational semantics. The denotational semantics provides the semantics for a modal logic of agent programs [11].

### B.  Statistical Evaluation

This evaluation is based on the classical statistical approach of hypothesis testing. The assumed hypothesis (also called Null hypothesis, $H_0$) is that the output data of the agent model follows certain perfect statistical distribution.

The SRdr agent acts accordingly to the distribution pattern of the number of targets detected ($n_t$). The cut off value of the NTD has a role in the performance measure of the SRdr agent. If the number of targets detected ($n_t$) follows the Gaussian random distribution then the *NTD* follows the *Student's- t* distribution. Some other form of transformations is given in the Table 1.

The decision of correct detection of jamming is a classical statistical problem of finding signal in the background of random noise. The Kolmogorov-Smirnov (*KS*) statistic is used to determine the underlying distribution pattern of *NTD* as given in the Figure 9. Probability of false alarm plays a vital role in the correct detection of jamming. Although a single value of false alarm (i.e. 5%) is taken in this study, the performance measure of the agent can be simulated for other values of $P_f$.

To test the SRdr agent performance, 500 random numbers were generated using the Gaussian random distribution with mean and standard deviation equal to 20 and 10 respectively. On the generated data, the *NTD* index is calculated. It is found that after *NTD* transformation, the Gaussian random number transforms into another form of statistical distribution which is very similar to the S*tudent's- t* distribution with parameter $v = 2$. If it is assumed that the $n_t$ follows some other statistical distribution then the resulting distributions of *NTD* would be of the form given in the Table 1.

The SRdr agent's algorithm is applied to the generated data. The result of the simulation after applying the SRdr agent's logic is shown in the Table 2. This table shows that out of total 500 samples, 231 times (i.e. 41 %) the radar is found to be jammed. This statistic is close to the *Jamming Factor* (which was introduced in the simulation as random noise). It was found that 91 times the radar is found "*Switch Off*". So it has saved around 19% of the energy. Although the simulation was started with Gaussian random data, after applying the agent's logic the data transformed into statistical distribution which is very similar to the Binomial distribution (because the transformed data, the decisions, were either "*Switch On*" or "*Switch Off*"). The closeness of this distribution is measured by the *KS* statistic as shown in the Figure 9. This statistic is used for performance measure of this agent. This performance measure could be used to add the learning capability in the agents.

For the LCCC agent, experiments have been performed with three clusters, three types of cluster size (i.e. large, medium and small) and two types of mission objectives (i.e. strike and escort). So, total eighteen (3 × 3 × 2) possibilities of plans instances are generated. Therefore, search space consists of eighteen combinations. Hence, it is obvious with the increase

- Goal　　: $K/\beta \Rightarrow \pi$ : if to achieve goal $k$ for given belief $\beta$ then perform the plan $\pi$.
- Goal Inference Rule (gir):
  $\{Jammed\}^{\beta}, \{Frequency\ Hopping\}^{k-} \Rightarrow Switch\ Off$
  $\{Jammed\}^{\beta}, \{Switch\ Off\}^{k-} \Rightarrow Frequency\ Hopping$
  $\{Frequency\ Hopping\}^{k+} \Rightarrow Sense\ Mode$
  $\{Switch\ Off\}^{k+} \Rightarrow Sleep\ Mode$
- Tautology　: $\neg$ *Frequency Hopping, Switch Off / Switch Off,*
- Tautology　: $\neg$ *Switch Off, Frequency Hopping / Frequency Hopping,*
  *Frequency Hopping*: *Sense Mode / Sense Mode,*
  *Switch Off*: *Sleep Mode / Sleep Mode.*
  Extension　: $\{Switch\ Off,\ Sleep\ Mode\}, \{Frequency\ Hopping,\ Sense\ Mode\}$
- Goal　　: G$\{Switch\ Off \wedge Sleep\ Mode\}$, G$\{Frequency\ Hopping \wedge Sense\ Mode\}$
- Conflicting Goal :
  $\neg$ G$\{Switch\ Off \wedge Frequency\ Hopping\}$,
  $\neg$ G$\{Sense\ Mode \wedge Sleep\ Mode\}$,
  $\neg$ G$\{Switch\ Off \wedge Sense\ Mode\}$,
  $\neg$ G$\{Frequency\ Hopping \wedge Sleep\ Mode\}$

**Figure 8. Goal inference rule (gir) [12] for logical verification of surveillance radar agent. G stands for goal.**

of search space computation time taken by the agent decreases. Similarly, the number of *VAVP* points also influences the agent's performance. Although in this study only the *VAVP* points are considered in the agent's beliefs, the number of interceptor also influences the agent's performance, therefore, can be included in the beliefset. In general, the number of domains of the input parameters determines the performance of the agents. Search space will increase multiplicatively with the increase of the domain size. Computation time to take an optimal decision using MLPR is effected by these factors.

The input from the MSDF module influences the computation time required by the LCCC agent. It is found that the number of cluster has direct influences on the performance of the LCCC agents. In this study, only three clusters are considered. How the agent will perform with many clusters have not been studied. Similarly when the number of mission type and cluster size changes it directly influences the ranking of the plan instance.

## DISCUSSION AND CONCLUSIONS

The SRdr and LCCC agents are deployed in a simulated environment of air combat. The simulation is designed with several entities like attacking aircrafts, defending interceptors, surveillance radars, air to air and surface to air missiles, tracking radars. The simulated environment is created by Java Netbeans IDE [18]. The agents are programmed by the JACK agent programming language [17]. JACK supports BDI architectures and MLPR. The data generated by this simulation are stored in the Oracle [19] database. The output actually contains the information about the states conditions of the environment and agents. The agents analyse the environment and write their decisions again in this database. The initial belief of the SRdr agent is that no jamming has occurred. Over the time this agent keeps on reading data from the database and adds it to its belief set which automatically posts an event if it is greater than 0.5 (a threshold value decided by the experts) as an indication of noise jamming. The MSDF module collaboratively assess the data obtain by different sensors and writes in the database. The LCCC agent receives inputs from the MSDF output through this database and decides accordingly. Based on these decisions, resources are allocated to the attacking aircrafts.

On each run of the simulation state situation of both the environment and the agents are observed. The agents are programmed such a way that these can automatically detect any conflicting goal situation. For example, the SRdr agent checks its present states and if it finds any conflicting states situation as shown in the Figure 8 it throws an exception. In this way the agent model is validated logically. For statistical performance evaluation, the *KS* statistic is used. The *KS* statistic with a lower value is always preferable.

The main point emphasized in this work is the implementation details. Also, the MLPR is introduced so agents can choose the right plan for the plan-repository using a prioritizing mechanism. Such reasoning and the way it is implemented can be used for many different application domains. Main contribution of the paper is combining agent-oriented programming with a Java based simulation environment and implementing this for an IAD domain. Main focus of the work is on how it is done and a significant effort has been put in implementing these ideas. Given the level of details it is certain that an advanced system may be developed for further research is this field.

The present approaches of design, implementation and testing of agent based system are found to be more suitable for hierarchical structure of C2 that works on the principle of practical reasoning. The way BDI architectures are used for developing the C2 agents can be extended to build higher order team agents. This could be a general frame work for implementing decision making processes in an integrated mode. The traditional optimization technique used for TA for air defense system can be brought into this framework very easily. This is an integrated approach of decision making for selecting the optimal plan satisfying the agent's beliefs to achieve desired goals. Usual methods of decision making do not integrate the decision maker's beliefs and desires in direct way although these components are essential attributes. This approach is more suitable for futuristic network centric warfare. The approach is novel in terms of both implementation (MLPR) and validation (logical as well as statistical).

| No. of Targets Detected with parameters | Normalized Target Difference |
|---|---|
| Normal (20,10) | Student's t (2) |
| Triangular (20,10,30) | Gamma ($\alpha$=12.06, $\beta$=0.08) |
| Uniform (10,30) | Gamma ($\alpha$=4.90, $\beta$=0.22) |
| Exponential (10,20) | Laplace ($\lambda$=185.71, $\mu$=1.0) |

**Table 1. Distribution pattern of Normalized Target Difference as obtained from the transformation of distribution of numbers of targets detected ($n_t$).**

| Events | Number/ Percentage |
|---|---|
| Total Samples | 500 |
| Jamming | 231(46.2%) |
| Frequency Hopping | 134(26.8%) |
| Switch Off | 97(19.4%) |

**Table 2. Simulation result of the surveillance radar agent.**

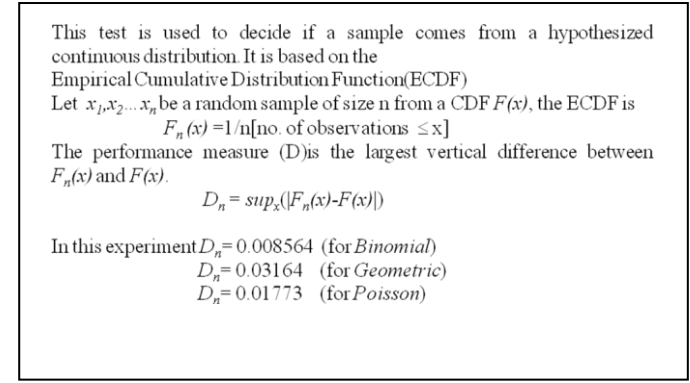

Figure 9. Kolmogorov Smirnov statistics for agent's performance measurement.

*Dr Sumanta Kumar Das and Mr. Sumant Mukherjee are scientists in the Institute for Systems Studies and Analyses, Defence Research and Development Organization, Ministry of Defence, Delhi, India. Both of them are working together for developing the agent based models for integrated air defense system especially for network centric warfare. E-mail: sumantadas.delhi@ieee.org.*